\documentclass[aps,singlecolumn,superscriptaddress,nofootinbib,tightenlines,
preprintnumbers,showkeys]{revtex4-1}

\usepackage[utf8]{inputenc}
\usepackage[english]{babel}
\usepackage{amssymb,amsthm,amsmath,amstext,amsbsy,amsopn}
\usepackage{bbm}
\usepackage{nicefrac}
\usepackage{slashed}
\usepackage{graphicx}
\usepackage{hyperref}
\usepackage{leftidx}
\usepackage{environ}
\usepackage{mathtools}
\usepackage{xspace}
\usepackage{array}

\newcommand{\eg}{\textit{e.g.}}
\newcommand{\cf}{\textit{cf.}\xspace}
\newcommand{\apriori}{\textit{a priori}\xspace}

\newcommand{\vNabla}{\boldsymbol{\nabla}}

\newcommand{\ee}{\mathrm{e}}

\newcommand{\bra}[1]{\langle #1|}

\newcommand{\ket}[1]{|#1\rangle}

\newcommand{\abs}[1]{\left|#1\right|}

\newcommand*\rvec[1]%
{\ensuremath{\overset{\smash{\raisebox{-1.5pt}{\tiny$\rightarrow$}}}{#1}}}
\newcommand*\lvec[1]%
{\ensuremath{\overset{\smash{\raisebox{-1.5pt}{\tiny$\leftarrow$}}}{#1}}}

\newcommand{\alatt}{\ensuremath{a_{\text{latt}}}}

\NewEnviron{subalign}[1][]{%
\begin{subequations}\begin{align}
  \BODY
\end{align}\label{#1}\end{subequations}
}

\NewEnviron{spliteq}{%
\begin{equation}\begin{split}
  \BODY
\end{split}\end{equation}
}

\newcolumntype{K}[1]{>{\centering\arraybackslash}p{#1}}

\newcommand{\ten}[1]{\times10^{#1}}

\begin{document}

\title{Volume Dependence of ${\boldsymbol N}$-Body Bound States}

\keywords{Finite-volume correction; bound states; lattice field theory}

\author{Sebastian König}
\email{sekoenig@theorie.ikp.physik.tu-darmstadt.de}
\affiliation{Institut für Kernphysik, Technische Universität Darmstadt, 
64289 Darmstadt, Germany}
\affiliation{ExtreMe Matter Institute EMMI,
GSI Helmholtzzentrum für Schwerionenforschung GmbH,
64291 Darmstadt, Germany}

\author{Dean Lee}
\email{leed@nscl.msu.edu}
\affiliation{National Superconducting Cyclotron Laboratory, 
Michigan State University, MI 48824, USA}
\affiliation{Department of Physics, North Carolina State University,
Raleigh, NC 27695, USA}

\date{\today}

\begin{abstract}
We derive the finite-volume correction to the binding energy of an
$N$-particle quantum bound state in a cubic periodic volume.  Our results are 
applicable to bound states with arbitrary composition and total angular 
momentum, and in any number of spatial dimensions.  The only assumptions are 
that the interactions have finite range.  The finite-volume correction is a sum 
of contributions from all possible breakup channels.  In the case where the 
separation is into two bound clusters, our result gives the leading volume 
dependence up to exponentially small corrections.  If the separation is into 
three or more clusters, there is a power-law factor that is beyond the scope of 
this work, however our result again determines the leading exponential 
dependence.  We also present two independent methods that use finite-volume data 
to determine asymptotic normalization coefficients.  The coefficients are useful 
to determine low-energy capture reactions into weakly bound states relevant for 
nuclear astrophysics.  Using the techniques introduced here, one can even 
extract the infinite-volume energy limit using data from a single-volume 
calculation.  The derived relations are tested using several exactly solvable 
systems and numerical examples.  We anticipate immediate applications to lattice 
calculations of hadronic, nuclear, and cold atomic systems.
\end{abstract}

\maketitle

\section{Introduction}

In a number of highly influential
papers~\cite{Luscher:1985dn,Luscher:1986pf,Luscher:1990ux}, Lüscher derived 
the volume dependence of two-particle bound states and scattering states in
cubic periodic volumes.  The bound-state relation connects the finite-volume 
correction to the asymptotic properties of the two-particle wave function,
whereas the elastic scattering result relates the volume dependence of discrete 
energy levels to physical scattering parameters.  This work has since been 
extended in several directions, including non-zero angular
momenta~\cite{Luu:2011ep,Konig:2011nz,Konig:2011ti}, moving
frames~\cite{Kim:2005gf,Rummukainen:1995vs,Bour:2011ef,Davoudi:2011md,
Rokash:2013xda}, generalized boundary
conditions~\cite{Sachrajda:2004mi,Briceno:2013hya,Korber:2015rce,
Cherman:2016vpt,Schuetrumpf:2016uuk}, particles with intrinsic
spin~\cite{Briceno:2014oea}, and perturbative Coulomb 
corrections~\cite{Beane:2014qha}.\footnote{In a different but related approach, 
two-nucleon scattering properties have been extracted by solving the system in 
an artificial harmonic trap~\cite{Luu:2010hw}, based on results obtained for 
cold atoms, where the trap is 
physical~\cite{Busch:1997aa,Yip:2008aa,Suzuki:2009aa}.}

With improved numerical techniques and computational resources enabling the
calculation of systems with an increasing number of constituents, understanding 
the volume dependence of more complex systems is of timely relevance.  Currently 
some results are available for three-particle systems, ranging from the general 
theory~\cite{Polejaeva:2012ut,Briceno:2012rv,Briceno:2016ffu} to explicit 
results for specific
systems~\cite{Kreuzer:2010ti,Kreuzer:2012sr,Kreuzer:2013oya,Meissner:2014dea}.  
In this letter, we derive the volume dependence of $N$-particle bound states 
with finite-range interactions in $d$ spatial dimensions and arbitrary total 
angular momentum.  We also use finite-volume energies to extract asymptotic 
normalization coefficients, which are useful in halo effective field theory 
calculations of low-energy reactions of relevance for nuclear 
astrophysics~\cite{Xu:1994zz,Capel:2013zka,Zhang:2014zsa,Hammer:2017tjm}.  The 
results presented here should have numerous and immediate applications for 
lattice QCD and lattice effective field theory calculations of nuclei.
Moreover, our results also apply to lattice simulations of cold atomic systems, 
as discussed for example in Refs.~\cite{Endres:2012cw,Bour:2012hn,Bour:2014bxa}.

When the separation is into two bound clusters, the leading correction is the 
same as the finite-volume correction for a two-particle system, where the 
clusters are treated as though they were fundamental particles.  While one may 
have guessed this result in the case where the $N$-particle system is a weakly 
bound system of two clusters, we show that this formula continues to hold 
even when the $N$-particle system is more strongly bound than one or more of 
the constituent clusters.\footnote{In the finite volume, all energy levels are 
discrete states.  We refer to individual levels as bound and 
continuum/scattering states, respectively, if their extrapolated infinite 
volume energy is below or above the non-interacting $N$-body threshold.  In the 
finite volume, bound states defined this way are characterized by an exponential 
dependence on the volume whereas continuum/scattering states have a power-law 
volume dependence.}

In the case where the separation is into three or more clusters, our derivation 
gives the leading exponential dependence.  However, in this case there are also 
correction factors which scale as inverse powers of the periodic box size.  We 
discuss these power-law factors here for a few special cases, while the general 
result will be addressed in a future publication.

\section{Asymptotic behavior}

We start with $N$ nonrelativistic particles in $d$ spatial dimensions with 
masses $m_1,\cdots m_N$.  We are using units where $\hbar=c=1$ and write the 
position-space wave function for a general state $\ket\psi$ as $\psi({\bf 
r}_1,\cdots{\bf r}_N)$.  The Hamiltonian we consider is of the form
\begin{equation}
 \hat{H}_{1\cdots N} = \sum_{i=1}^N \hat{K}_{i}+\hat{V}_{1\cdots N} \,,
\end{equation}
where $\hat{K}_{i} = {-\vNabla}^2_i/{(2m_i)}$, and in general we have nonlocal 
interactions of every kind from two-particle up to $N$-particle interactions.
We can write the total interaction as a sum of integral kernels,
\begin{equation}
 V_{1\cdots N}({\bf r}_1,\cdots{\bf r}_N ;{\bf r}'_1,\cdots{\bf r}'_N)
 = \sum_{i < j} W_{i,j}({\bf r}_i,{\bf r}_j;
 {\bf r}'_i,{\bf r}'_j)1_{\slashed{i},\slashed{j}}
 \null + \sum_{i< j< k}W_{i,j,k}({\bf r}_i,{\bf r}_j,{\bf r}_k;
 {\bf r}'_i,{\bf r}'_j,{\bf r}'_k)1_{\slashed{i},\slashed{j},\slashed{k}}
 + \cdots \,,
\label{eq:vtot}
\end{equation}
where we use the shorthand notation
\begin{equation}
 1_{\slashed{i}_1,\cdots\slashed{i}_k}
 = \prod_{j\ne{{i}_1,\cdots {i}_k}}\delta^d({\bf r}_j-{\bf r}'_j)
\end{equation}
for the spectator particles.  We assume that the interactions respect Galilean 
invariance, and so the center-of-mass (c.m.) momentum is conserved, and the 
c.m.\ kinetic energy decouples from the relative motion of the $N$-particle 
system.  We furthermore assume that every interaction has finite range, meaning 
that each $W_{i_1\cdots i_k}$ vanishes whenever the separation between some pair 
of incoming or outgoing coordinates exceeds some finite length $R$.

We now consider an $N$-particle bound state with total c.m.\ momentum zero,
binding energy $B_N$, and wave function $\psi^B_N({\bf r}_1,\cdots{\bf r}_N)$.  
In our notation the binding energy is the absolute value of the bound-state 
energy.  Let us consider the asymptotic properties of this wave function when 
one of the coordinates becomes asymptotically large, while keeping the others 
fixed.  Without loss of generality, we take the coordinate that we pull to 
infinity to be ${\bf r}_1$.

Let $S$ refer to the set of coordinate points $\{ {\bf r}_1,\cdots{\bf r}_N
\}$ where ${\bf r}_1$ is greater than distance $R$ from all other coordinates.
Therefore in $S$ there are no interactions coupling ${\bf r}_1$ to
${\bf r}_2,\cdots{\bf r}_N$.  By the assumption of vanishing c.m.\ momentum, we 
can work with the reduced Hamiltonian
\begin{equation}
 \sum_{i=2}^{N} \hat{K}_i - \hat{K}^{\rm CM}_{2\cdots N}
 + \hat{V}_{2\cdots N} + \hat{K}^{\rm rel}_{1|N-1} \,,
\label{eq:reduced}
\end{equation}
where $\hat{K}^{\rm CM}_{2\cdots N}
= -({\vNabla}_2+\cdots{\vNabla}_{N})^2/(2m_{2 \cdots N})$ and
\begin{equation}
\hat{K}^{\rm rel}_{1|N-1}
 = {-}\frac{\left(m_{2\cdots N}{\vNabla}_1 - m_1{\vNabla}_{2\cdots N}\right)^2}
 {2\mu_{1|N-1}m^2_{1\cdots N}} \,.
\end{equation}
We have written $m_{n\cdots N} = m_n+\cdots+m_{N}$ for the total mass of the 
(sub)system for the two cases $n=1$ and $n=2$.  We have also introduced 
$\mu_{1|N-1}$ as the reduced mass with $\tfrac{1}{\mu_{1|N-1}} = 
\tfrac{1}{m_1}+\tfrac{1}{m_{2\cdots N}}$.

We note that the first three terms in Eq.~(\ref{eq:reduced}) constitute the 
Hamiltonian $\hat{H}_{2\cdots N}$ of the ${\{2,\cdots N\}}$ subsystem with the 
c.m.\ kinetic energy removed, while the remaining $\hat{K}^{\rm rel}_{1|N-1}$ 
is the kinetic energy of the relative motion between particle $1$ and the
center of mass of the ${\{2,\cdots N\}}$ subsystem.  In region $S$ we use the 
separation of variables to expand $\psi^B_N({\bf r}_1,\cdots {\bf r}_N)$ as a 
linear combination of products of eigenstates of $\hat{H}_{2\cdots N}$ with 
total linear momentum zero and eigenstates of $\hat{K}^{\rm rel}_{1|N-1}$.

For the moment we assume that the ground state of $\hat{H}_{2\cdots N}$ is a
bound state with energy ${-}B_{N-1}$ and wave function
$\psi^B_{N-1}({\bf r}_2,\cdots{\bf r}_{N})$.  For simplicity we consider here
the case where the relative motion wave function has zero orbital angular 
momentum and will relax this condition later in the discussion.  Then, as 
$r_{1|N-1}=\abs{{\bf r}_{1|N-1}}$ becomes large, we have
\begin{equation}
 \psi^B_N({\bf r}_1,\cdots{\bf r}_N)
 \propto \psi^B_{N-1}({\bf r}_2,\cdots{\bf r}_{N})
 \null (\kappa_{1|N-1}r_{1|N-1})^{1-d/2} \,
 K_{d/2-1}(\kappa_{1|N-1}r_{1|N-1}),
\label{eq:asymptotic1}
\end{equation}
where $K_{d/2-1}$ is a modified Bessel function of the second kind,
${\bf r}_{1|N-1} = {\bf r}_1
- (m_2{\bf r}_2+\cdots+m_{N}{\bf r}_{N})/{m_{2\cdots N}}$, and
\begin{equation}
 \kappa_{1|N-1} = \sqrt{2\mu_{1|N-1}(B_{N}-B_{N-1})} \,.
\label{eq:1|N-1}
\end{equation}
For the excited states of the $N{-}1$ system there will be terms analogous
to Eq.~(\ref{eq:asymptotic1}), however they will be exponentially suppressed 
due to the larger energy difference with $B_{N}$.

This discussion is readily generalized to the case of two clusters with 
arbitrary particle content.  For this case we take the center of mass of $A$ 
coordinates to infinity while keeping the relative separations within the $A$ 
and $N{-}A$ subsystems fixed.  Without loss of generality, we can choose the 
$A$ coordinates to be ${\bf r}_1,\cdots{\bf r}_{A}$.  Following steps analogous 
to the case $A=1$, we again apply the separation of variables to the
$N$-particle wave function and obtain
\begin{equation}
 \psi^B_N ({\bf r}_1,\cdots{\bf r}_N)
 \propto \psi^B_{A}({\bf r}_1,\cdots{\bf r}_{A})
 \psi^B_{N-A}({\bf r}_{A+1},\cdots{\bf r}_{N})
 \null (\kappa_{ A|N-A} r_{A|N-A})^{1-d/2} \,
 K_{d/2-1}(\kappa_{A|N-A}r_{A|N-A}) \,,
\label{eq:asymptotic2}
\end{equation}
where
\begin{align}
 {\bf r}_{A|N-A}
 &= \tfrac{m_1{\bf r}_1+\cdots+ m_{A}{\bf r}_{A}}{m_1+\cdots+m_A}
 -\tfrac{m_{A+1}{\bf r}_{A+1}
 + \cdots + m_{N}{\bf r}_{N}}{m_{A+1}+\cdots+m_N} \,, \\
 \tfrac{1}{\mu_{A|N-A}}
 &= \tfrac{1}{m_1+\cdots+ m_A}+\tfrac{1}{m_{A+1}+\cdots+m_N} \,, \\
 \kappa_{A|N-A}
 &= \sqrt{2\mu_{A|N-A}(B_{N}-B_{A}-B_{N-A})} \,,
\end{align}
and ${-}B_{A}$ and ${-}B_{N-A}$ are the ground state energies of the 
$A$-particle and $(N{-}A)$-particle systems respectively.  We have made the 
simplifying assumption that $-B_{A}$ and $-B_{N-A}$ are both bound-state 
energies.  If this is not true and one or both are instead energies associated 
with a scattering threshold, then Eq.~(\ref{eq:asymptotic2}) remains correct up 
to additional prefactors that scale as inverse powers of 
$\kappa_{A|N-A}r_{A|N-A}$.  These factors arise from the integration over 
scattering states, and will be discussed in a future publication.

We now remove the condition that the relative motion between clusters have zero 
orbital angular momentum.  In the general case, the relative-motion wave
function has the form
\begin{equation}
 \sqrt{\tfrac{2\kappa_{ A|N-A}}{\pi}}\,r_{A|N-A}^{1-d/2}\sum_{{\bf\ L}}
 \gamma_{\bf{L}} 
 Y_{\bf{L}}(\hat{{\bf r}}_{A|N-A})
 \null K_{\ell+d/2-1}(\kappa_{A|N-A}r_{A|N-A}) \,,
\label{eq:ANC1}
\end{equation}
where $Y_{\bf{L}}$ denotes the $d$-dimensional hyperspherical harmonics for
spin representation $\ell$ (see for example Ref.~\cite{Hammer:2010fw}) and 
$\gamma_{\bf{L}}$ are constant coefficients.  This is exactly the same behavior
as found in two-particle bound states with nonzero angular momentum.  All of 
the various cases for $d=2$ and $d=3$ are discussed in Ref.~\cite{Konig:2011ti}.
For the one-dimensional case, $\ell={\bf L}=0$ corresponds with even parity and 
$\ell ={\bf L}=1$ corresponds with odd parity.  For even parity the spherical 
harmonic is just unity, while for odd parity it is an odd step function.

\section{Finite volume correction}

We define $B_N(L)$ as the finite-volume binding energy in a cubic periodic box
of length $L$ and let $\Delta B_N(L) = B_N(L) - B_N$ be the finite-volume 
correction.  Following steps analogous to 
Refs.~\cite{Luscher:1985dn,Konig:2011nz,Konig:2011ti,Meissner:2014dea}, we find 
that $\Delta B_N(L)$ gets contributions from every possible breakup channel.

If the $N$-particle system can be subdivided as an $A$-particle bound state and
$(N{-}A)$-particle bound state in a relative $\ell = 0$ state, then we get a 
contribution to $\Delta B_N(L)$ that is proportional to
\begin{equation}
 (\kappa_{ A|N-A}L)^{1-d/2} \, K_{d/2-1}(\kappa_{ A|N-A}L) \,.
\label{eq:L_dependence}
\end{equation}  
In addition to this there are also terms that have a larger exponential
suppression and thus can be neglected.  While generally different subdivisions 
contribute to the overall volume dependence of a given state, the smallest value 
$\kappa_{A|N{-}A}$ determines the leading asymptotic behavior.

If the two bound states have orbital angular momentum $\ell > 0$, then the 
finite volume correction has the same dependence as in 
Eq.~(\ref{eq:L_dependence}) along with subleading terms that are suppressed by 
powers of $\kappa_{A|N-A}L$.  The functional form is exactly the same as that 
for the $N=2$, $d=2,3$ cases derived in Refs.~\cite{Konig:2011nz,Konig:2011ti}. 
The sign of the correction oscillates with even and odd $\ell$.

For the case that the $A$-particle ground state, $(N{-}A)$-particle ground 
state, or both ground states are scattering states, we still obtain the same 
exponential dependence, except there is an additional power law factor of
$P(\kappa_{A|N-A}L)$ due to the integration over scattering states,
\begin{equation}
 (\kappa_{ A|N-A}L)^{1-d/2} K_{d/2-1}(\kappa_{ A|N-A}L)P(\kappa_{A|N-A}L) \,.
\label{eq:L_dependence_continuum}
\end{equation}
The general functional form for this power law factor $P(\kappa_{A|N-A}L)$ is 
beyond the scope of this current study, but we discuss the power-law factor 
$P(\kappa_{ A|N-A}L)$ for several notable examples below.  The remaining 
finite-volume corrections are exponentially suppressed compared to the terms in 
Eq.~(\ref{eq:L_dependence}) and Eq.~(\ref{eq:L_dependence_continuum}).

\section{Analytically solvable examples}

We now consider several examples to check the results we have derived.
We start with the exactly solvable $N$-particle system in one dimension with all 
equal masses $m$ and an attractive delta-function potential ${-}c\delta(r_i - 
r_j)$ between every pair of particles $i,j$.  The exact ground state binding 
energies are
\begin{equation}
 B_N = \frac{N(N^2-1)}{6}B_2 \,,
\end{equation}
where $B_2=c^2m/4$, and the exact wave functions are
\begin{equation}
 \psi^B_N (r_1,\cdots r_N)
 \propto \exp\!\left[-\kappa\sum_{i>j}|r_i-r_j|\right] \,,
\end{equation}
with $\kappa = cm/2$.  Let us now pull $r_1,\cdots r_A$ in unison to infinity 
with $r_{A+1},\cdots r_N$ fixed.  From the exact wave function we have
\begin{equation}
 \psi^B_N (r_1,\cdots r_N) \propto \psi^B_{A} (r_1,\cdots r_A) \, \psi^B_{N-A}
 (r_{A+1},\cdots r_N)
 \null \ee^{-A(N-A)\kappa|r_{A|N-A}|} \,.
\end{equation}
This result agrees precisely with the asymptotic behavior given in 
Eq.~\eqref{eq:asymptotic2}.

\medskip
The next example we consider is the three-particle system at the unitarity
limit in three dimensions.  This corresponds with two-particle interactions with
zero range and infinite scattering length.  Our result for the leading 
finite-volume correction is
\begin{equation}
 \Delta B_3(L) \propto (\kappa_{1|2}L)^{-1/2} 
 K_{1/2}(\kappa_{1|2}L)P(\kappa_{1|2}L)
 \propto\exp\!\left({-}\sqrt{\tfrac{4mB_3}{3}}L\right)
 \left({\sqrt{\tfrac{4mB_3}{3}}L}\right)^{\!{-}1}P(\kappa_{1|2}L) \,,
\end{equation}
where $P(\kappa_{1|2}L)$ is a power-law function arising from the scattering
threshold of the two-particle system.  We find the same exponential dependence
as the result derived in Ref.~\cite{Meissner:2014dea},
\begin{equation}
 \Delta B_3(L) \propto 
 \exp\!\left({-}\sqrt{\tfrac{4mB_3}{3}}L\right)\left({\sqrt{\tfrac{4mB_3}{3}}L}
\right)^{\!{-}3/2}+\cdots \,,
\end{equation}
and there are subleading power-law corrections in addition to exponentially
suppressed corrections.  In this case we find that the leading power-law
behavior is $P(\kappa_{1|2}L)=(\kappa_{1|2}L)^{-1/2}$.

\medskip
Finally, the third example we consider is the case of a spinless $N$-particle 
bound state with only an $N$-particle interaction.  In this case there are no 
cluster substructures and we need only consider the asymptotic behavior as one 
of the particles is pulled away from the center of mass of the remaining 
ones.  Without loss of generality, let the particle that we pull away be ${\bf 
r}_1$.  The corresponding Jacobi coordinate is ${\bf r}_{1|N-1}$ with reduced 
mass $\mu_{1|N-1}$.  As we pull ${\bf r}_1$ away from the center of mass of the 
other particles, the wave function can be shown to be proportional to
\begin{equation}
 (\kappa_{1|N-1}r_{1|N-1})^{1-d(N-1)/2} \,
 K_{d(N-1)/2-1}(\kappa_{1|N-1}r_{1|N-1}) \,.
\end{equation}
This result matches the same exponential dependence as our predicted result
\begin{equation}
 (\kappa_{1|N-1}r_{1|N-1})^{1-d/2} \, K_{d/2-1}(\kappa_{1|N-1}r_{1|N-1})
 \null  P(\kappa_{1|N-1}r_{1|N-1}) \,.
\end{equation}
We can further conclude that in this case the power law suppression factor is
\begin{equation}
 P(\kappa_{1|N-1}r_{1|N-1}) \propto (\kappa_{1|N-1}r_{1|N-1})^{-d(N-2)/2}
\,.
\end{equation}

\section{Numerical Tests}

We now test our results with numerical calculations.  The program used to 
calculate the results presented in this section has been set up to handle an 
arbitrary number of particles and spatial dimensions, limited only by available 
computational resources.  This is achieved through a generator code (written in 
Haskell) that automatically creates scripts (to be run with GNU Octave or 
compatible software) for each desired case.  This Haskell code is provided as 
supplementary material together with this letter (see appendix for details).

We consider equal-mass particles interacting via attractive local Gaussian 
potentials\footnote{While these potentials do not have a strictly finite range, 
their fall-off at large distances is much faster than any expected volume 
dependence, such that our relations hold up to negligibly small corrections.  
The use of Gaussian potentials instead of, \eg, strictly finite-range step 
potentials has the advantage of avoiding large discretization artifacts.} in 
$d=1,2,3$ dimensions.  For the derivatives appearing in the kinetic energy, we 
write the finite difference up to $k$-th order accuracy, where $k=2,4,\cdots$.

Expanding the Bessel function in Eq.~\eqref{eq:L_dependence}, we find that
the leading finite-volume correction has the asymptotic form
\begin{equation}
 \Delta B_N(L) \propto \exp\left({-}\kappa_{A|N-A}L\right) / L^{(d-1)/2} \,.
\label{eq:DeltaB-ND}
\end{equation}
This form can be easily identified by plotting the logarithm of $\Delta B_N(L)$
times $L^{(d-1)/2}$ as a function of $L$, and linear fits can be used to
extract the slopes to be compared to the expected $\kappa_{A|N{-}A}$.
%
\begin{table}[htbp]
\centering
\begin{tabular}{cccll}
\hline\hline
$\rule{0pt}{1.2em}\phantom{x}N\phantom{x}$
& $B_N$
& $L_{\text{min}} \ldots L_{\text{max}}$
& $\hspace{1.5em}\kappa_{\text{fit}}$ 
& $\kappa_{1|N-1}$ \\
\hline\hline
\multicolumn{5}{c}{\rule{0pt}{1.2em}
$d=1$, $V_0 = {-}1.0$, $R = 1.0$} \\
\hline
\rule{0pt}{1.2em}%
2 & 0.356 & $20\ldots48$ & $0.59536(3)$ & 0.59625 \\
3 & 1.275 & $15\ldots32$ & $1.1062(14)$ & 1.1070 \\
4 & 2.859 & $12\ldots24$ & $1.539(3)$   & 1.541 \\
5 & 5.163 & $12\ldots20$ & $1.916(21)$  & 1.920 \\
\hline\hline
\multicolumn{5}{c}{\rule{0pt}{1.2em}
$d=2$, $V_0 = {-}1.5$, $R = 1.5$} \\
\hline
\rule{0pt}{1.2em}%
2 & 0.338 & $15\ldots36$ & $0.58195(6)$ & 0.58140 \\
3 & 1.424 & $12\ldots24$ & $1.20409(3)$ & 1.20339 \\
4 & 3.449 & $7\ldots14$  & $1.743(8)$    & 1.743 \\
\hline\hline
\multicolumn{5}{c}{\rule{0pt}{1.2em}
$d=3$, $V_0 = {-}5.0$, $R=1.0$} \\
\hline
\rule{0pt}{1.2em}%
2 & 0.449 & $15\ldots24$ & $0.6694(2)$ & 0.6700 \\
3 & 2.916 & $4\ldots14$ & $1.798(3)$ & 1.814 \\
\hline\hline
\end{tabular}
\caption{Numerical results for local Gaussian potentials $V(r) = 
V_0\exp(-r^2/R^2)$.  All quantities are given in units of the particle mass 
$m=1$ (see text).}
\label{tab:Results-Gauss}
\end{table}

\begin{figure}[htb]
\centering
\includegraphics[width=0.6\columnwidth]{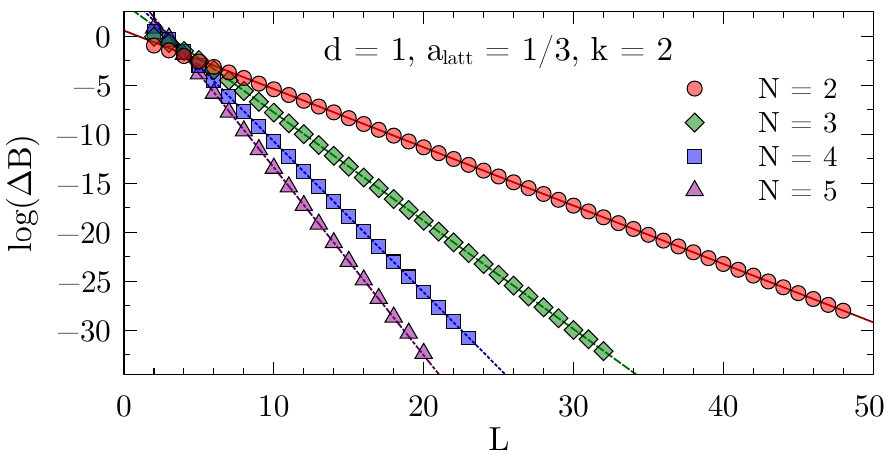}
\caption{(Color online) Finite-volume energy shift for $N=2,3,4,5$ particles 
interacting via a Gaussian potential ($R=1$, $V_0=-1$) in one dimension.  All 
quantities are given in units of the particle mass $m=1$ (see text).}
\label{fig:En-1D-Gauss}
\end{figure}

\begin{figure}[htb]
\centering
\includegraphics[width=0.6\columnwidth]{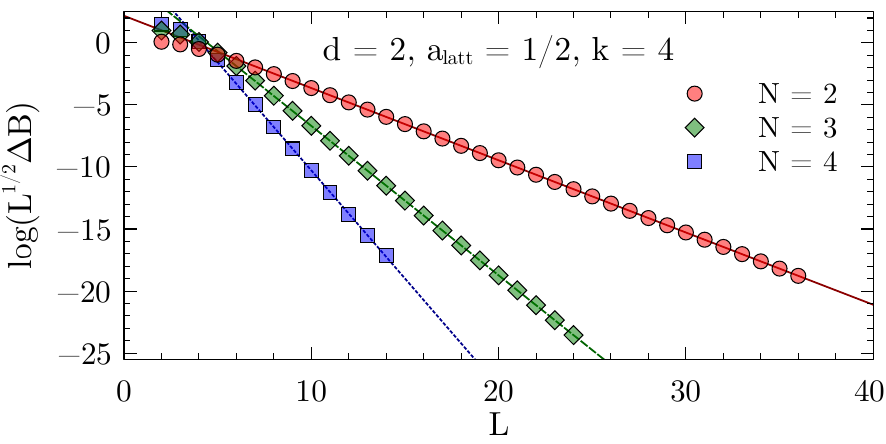}
\caption{(Color online) Finite-volume energy shift for $N=2,3,4$ particles 
interacting via a Gaussian potential ($R=1.5$, $V_0=-1.5$) in two dimensions.  
All quantities are given in units of the particle mass $m=1$ (see text).}
\label{fig:En-2D-Gauss}
\end{figure}

\begin{figure}[htb]
\centering
\includegraphics[width=0.6\columnwidth]{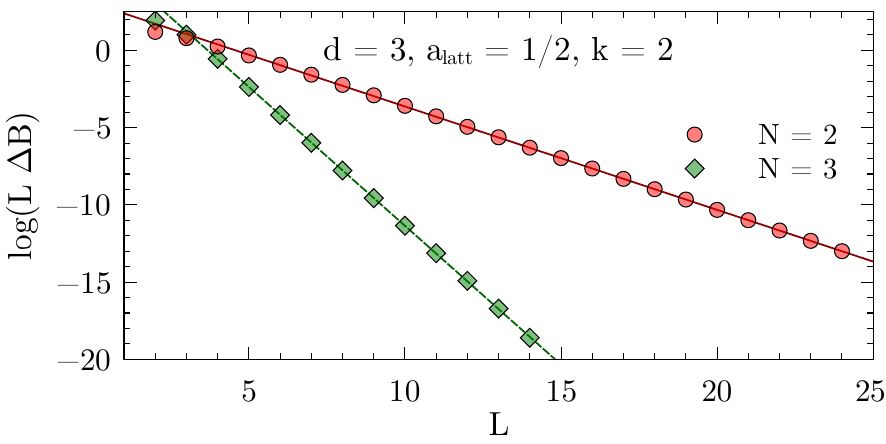}
\caption{(Color online) Finite-volume energy shift for $N=2,3$ particles 
interacting via a Gaussian potential ($R=1$, $V_0=-5$) in three dimensions.
All quantities are given in units of the particle mass $m=1$ (see text).}
\label{fig:En-3D-Gauss}
\end{figure}

Results for different potentials and spatial dimensions $d=1,2,3$ and are 
shown in Figs.~\ref{fig:En-1D-Gauss}, \ref{fig:En-2D-Gauss}, 
and~\ref{fig:En-3D-Gauss}.  In these calculations, all physical quantities are 
expressed in terms of powers of the particle mass $m$, which has been set to 
unity.  That is, energies and momenta are divided by the mass, and length scales 
are multiplied by the mass, and overall we set $\hbar=c=m=1$.  The lattice 
spacings $\alatt$ for the calculations were chosen to minimize discretization 
artifacts as much as possible while probing volumes large enough to test the 
asymptotic behavior of the finite-volume corrections.
%
%
From the observed straight lines in the plots it is clear that 
Eq.~\eqref{eq:DeltaB-ND} holds as a very good approximation.  The true 
infinite-volume energies $B_N$ used to calculate the shifts $\Delta B_N(L)$ were 
obtained by minimizing the error in fitting linear functions to the data points.

As summarized in Table~\ref{tab:Results-Gauss}, the slopes extracted from these
linear fits (performed within the ranges $L_{\text{min}} \ldots 
L_{\text{max}}$) agree very well with the prediction that the 
asymptotic volume dependence is dominated by the $1|N{-}1$ contribution.
These results provide further assurance that the relations derived in this 
letter are correct.  Similar agreement is found for other setups (involving 
higher-body potentials and/or relevant contributions other than $1|(N{-}1)$) as 
well; these will be presented in a forthcoming publication.

\section*{Asymptotic normalization coefficients}

Let us now consider the case where both the $A$-particle and $(N{-}A)$-particle
clusters are bound states or fundamental particles. In this case we extract an 
asymptotic normalization coefficient (ANC) associated with the $A+(N{-}A)$ 
threshold.  ANCs play an important role for low-energy capture processes that 
govern nucleosynthesis in stellar 
environments~\cite{Xu:1994zz,Capel:2013zka,Zhang:2014zsa,Hammer:2017tjm} and 
are notoriously difficult to extract in terrestrial experiments due to dominance 
of the Coulomb repulsion at low energies.  We will describe the problem of 
Coulomb ANCs in a future publication and focus here on the case of extracting
ANCs for finite-range interactions from finite-volume data.  

In the limit that separation distance $r_{A|N{-}A}$ between the two clusters is
large, the normalized $N$-body wave function is a product of normalized $A$-body 
and $(N{-}A)$-body wave functions times the relative wave function as written 
in Eq.~(\ref{eq:ANC1}).  The ANC is then the coefficient $\gamma_{\bf L}$ in 
Eq.~(\ref{eq:ANC1}), which we will write as $\gamma$ throughout the remainder 
of our discussion.  We note for $d=2$ our definition here differs from that used 
in Ref.~\cite{Konig:2011ti}.  We can extract the relative wave function by 
calculating the ratio
\begin{equation}
 \left(\frac{\bra {\Psi^B_N} O_A({\bf r}_{A|N-A}) O_{N-A}({\bf 0}) \ket 
 {\Psi^B_N}}
 {\bra {\Psi^B_A} O_A({\bf 0}) \ket {\Psi^B_{A}} \bra {\Psi^B_{N-A}}
 O_{N-A}({\bf 0}) \ket {\Psi^B_{N-A}}}\right)^{\!1/2}
\label{eq:Psi-O}
\end{equation}
for some localized $A$-body and $(N{-}A)$-body operators $O_{A}({\bf r})$, 
$O_{N-A}({\bf r})$.  We then compare this relative wave function at 
finite volume with the asymptotic form written in Eq.~(\ref{eq:ANC1}) plus 
additional copies due to the periodic boundary conditions, extracting the 
magnitude of the ANC.  We write this as $\abs{\gamma}_{\text{WF}}$, where WF 
is shorthand for wave function.

We can also determine the ANC in a completely different way using the 
finite-volume correction $\Delta B_N(L)$.  By combining our $N$-body results 
here with the derivations in 
Refs.~\cite{Luscher:1985dn,Konig:2011nz,Konig:2011ti}, we find that $\Delta 
B_N(L)$ equals
\begin{equation}
 \frac{(-1)^{\ell+1} \sqrt{\tfrac{2}{\pi}}
 f(d) \abs{\gamma}^2}{\mu_{A|N-A}}
 \kappa^{2-d/2}_{A|N-A}L^{1-d/2} K_{d/2-1}(\kappa_{A|N-A}L),
\label{eq:ANC-FV}
\end{equation}
plus smaller corrections that are exponentially suppressed.  The function $f(d)$
takes values $f(1)=2$, $f(2)=\sqrt{8/\pi}$, and $f(3)=3$.  If there are several
different ways to partition the $N$-particle system into clusters with same 
$\kappa_{A|N{-}A}$ value, then there will be contributions to the finite-volume 
correction from each channel.  In particular, in the case of $N$ identical 
particles (as considered in the numerical examples presented here), there is a 
combinatorial factor that counts the number of ways to partition the identical 
particles into $A|N{-}A$ clusters.  For $d=3$, $\Delta B_N(L)$ is averaged over 
all $2\ell +1$ elements of the angular momentum $\ell$ multiplet, while for 
$d=2$ the average is taken over symmetric and antisymmetric combinations of 
${\bf L} = \pm\ell$ for even $\ell$~\cite{Konig:2011ti}.

Equation~\eqref{eq:ANC-FV} follows directly from defining the ANC in terms of 
the asymptotic radial wave function, which for cluster separation $r_{A|N-A}$ 
large compared to the range of the interaction is universally given by
\begin{equation}
 \psi_{\text{asympt}}(r_{A|N-A})
 = \gamma \, \sqrt{\frac{2\kappa_{A|N-A}}{\pi}}
 (r_{A|N-A})^{1-d/2}
 \null K_{d/2-1}(\kappa_{A|N-A}r_{A|N-A}) \, Y(d) \,,
\label{eq:ANC-psi}
\end{equation}
where $Y(d)$ accounts for the angular normalization in $d$ spatial dimensions, 
\cf~Eq.~\eqref{eq:ANC1} specialized for the case of a bound state without 
angular dependence.  For $d=3$, where $Y(3) = 1/\sqrt{4\pi}$, the convention in 
Eq.~\eqref{eq:ANC-psi} reproduces the canonical form 
$\gamma\exp({-}\kappa_{A|N-A}r_{A|N-A})$ $/r_{A|N-A}$ for a two-cluster S-wave 
state.  For $d=1$ one has $Y(1) = 1$ and the asymptotic form is simply 
$\gamma\exp({-}\kappa_{A|N-A}r_{A|N-A})$.  For $d=2$, on the other hand, it is 
more natural to define the ANC directly in terms of the modified Bessel 
function, which does not reduce to a simple exponential in this case.  The 
$f(d)$ quoted above have been chosen to account for this as well as the
averaging over states in a cubic-group multiplet.  We write the magnitude of the 
ANC extracted from fits using Eq.~(\ref{eq:ANC-FV}) as 
$\abs{\gamma}_{\text{FV}}$, where FV is shorthand for finite volume.

Using the same lattice examples with Gaussian potentials as discussed 
previously, we present results for $\abs{\gamma}_{\text{FV}}$ and 
$\abs{\gamma}_{\text{WF}}$ in Table~\ref{tab:Results-Gauss-ANC}.  We use 
Eq.~\eqref{eq:Psi-O} with the operator $O_1$ equal to the single particle 
density and $O_{N-1}$ equal to the $(N{-}1)$-body density on a single lattice 
site, with all quantities extracted at the same finite volume.  As seen in 
Table~\ref{tab:Results-Gauss-ANC}, the two methods for extracting the ANCs are 
in excellent agreement.
%
\begin{table}[htbp]
\centering
\begin{tabular}{ccccc}
\hline\hline
$\rule{0pt}{1.2em}\phantom{x}N\phantom{x}$
& $B_N$
& $L_{\text{max}}$
& $\abs{\gamma}_{\text{FV}}$
& $\abs{\gamma}_{\text{WF}}$ \\
\hline\hline
\multicolumn{5}{c}{\rule{0pt}{1.2em}
$d=1$, $V_0 = {-}1.0$, $R = 1.0$} \\
\hline
\rule{0pt}{1.2em}%
2 & 0.356 & $48$ & $0.8652(4)$ & $0.8627(4)$ \\
3 & 1.275 & $32$ & $1.650(27)$ & $1.638(16)$ \\
4 & 2.859 & $24$ & $2.54(6)$   & $2.56(8) $ \\
5 & 5.163 & $20$ & $3.65(62)$  & $3.63(18)$ \\
\hline\hline
\multicolumn{5}{c}{\rule{0pt}{1.2em}
$d=2$, $V_0 = {-}1.5$, $R = 1.5$} \\
\hline
\rule{0pt}{1.2em}%
2 & 0.338 & $36$ & $1.923(2)$ & $1.921(9)$ \\
3 & 1.424 & $24$ & $5.204(4)$ & $5.24(2)$ \\
4 & 3.449 & $14$ & $11.2(4)$  & $10.99(4)$ \\
\hline\hline
\multicolumn{5}{c}{\rule{0pt}{1.2em}
$d=3$, $V_0 = {-}5.0$, $R=1.0$} \\
\hline
\rule{0pt}{1.2em}%
2 & 0.449 & $24$ & $1.891(3)$ & $1.89(1)$ \\
3 & 2.916 & $14$ & $7.459(97)$ & $7.83(11)$ \\
\hline\hline
\end{tabular}
\caption{Extracted ANCs for local Gaussian potentials $V(r) = 
V_0\exp(-r^2/R^2)$.  All quantities are given in units of the particle mass
$m=1$.}
\label{tab:Results-Gauss-ANC}
\end{table}

We furthermore note that by using the finite-volume wave function to determine 
$\abs{\gamma}_{\text{WF}}$ and $\kappa_{A|N{-}A}$, we can estimate $\Delta 
B_N(L)$.  By combining this with the finite-volume binding energy $B_N(L)$, we 
can determine the infinite-volume binding energy from a single-volume 
calculation.  We expect this technique to be of practical use for 
computationally expensive calculations of quantum bound states, perhaps making 
it unnecessary to perform infinite-volume extrapolations.  As an illustration, 
we show in Table~\ref{tab:Results-Gauss-DeltaB} results obtained by applying 
the method to the same potentials and states considered previously.  
Specifically, we calculate $\Delta B_N(L)_{\text{estimate}}$ by using 
$\abs{\gamma}_{\text{WF}}$ and $\kappa_{A|N{-}A}$ obtained from the wave 
function at a fixed volume $L=8$.  The uncertainties quoted are based on
varying the wave function tail range from which $\abs{\gamma}_{\text{WF}}$ and 
$\kappa_{A|N{-}A}$ are extracted using a fit to the known asymptotic form, and 
also by comparing the $\kappa_{A|N{-}A}$ thus determined to the result obtained 
directly from the energy at volume $L$.\footnote{For a fixed tail range, the 
ANC is determined from the minimum and maximum values of the ratio of the 
numerical wave function compared to the asymptotic form, while 
$\kappa_{A|N{-}A}$ is treated as a fit parameter.  The same technique 
was used to extract the $\abs{\gamma}_{\text{WF}}$ shown in 
Table~\ref{tab:Results-Gauss-ANC}, with the difference that this analysis used  
the $\kappa_{A|N{-}A}$ from the multiple-volume fit.}  Given that this error 
estimate does not include discretization artifacts or account for the subleading 
volume dependence of the states, we find overall good agreement for the 
single-volume extrapolation with the exact finite-volume corrections.
%
\begin{table}[htbp]
\centering
\begin{tabular}{cccrc}
\hline\hline
$\rule{0pt}{1.2em}\phantom{x}N\phantom{x}$
& $B_N$
& $\phantom{x}L\phantom{x}$
& $\Delta B_N(L)_{\text{estimate}}\phantom{x}$
& $\Delta B_N(L)_{\text{actual}}$ \\
\hline\hline
\multicolumn{5}{c}{\rule{0pt}{1.2em}
$d=1$, $V_0 = {-}1.0$, $R = 1.0$} \\
\hline
\rule{0pt}{1.2em}%
2 & 0.356 & $8$ & ${-}1.32(2)\ten{-2}$ & ${-}1.42\ten{-2}$ \\
3 & 1.275 & $8$ & ${-}3.9(4)\ten{-3}$ & ${-}3.75\ten{-3}$ \\
4 & 2.859 & $8$ & ${-}4.3(7)\ten{-4}$ & ${-}4.69\ten{-4}$ \\
5 & 5.163 & $8$ & ${-}0.6(2)\ten{-4}$ & ${-}0.64\ten{-4}$ \\
\hline\hline
\multicolumn{5}{c}{\rule{0pt}{1.2em}
$d=2$, $V_0 = {-}1.5$, $R = 1.5$} \\
\hline
\rule{0pt}{1.2em}%
2 & 0.338 & $8$ & ${-}2.5(6)\ten{-2}$ & ${-}2.84\ten{-2}$ \\
3 & 1.424 & $8$ & ${-}5.8(6)\ten{-3}$ & ${-}4.99\ten{-3}$ \\
4 & 3.449 & $8$ & ${-}4.1(6)\ten{-4}$ & ${-}4.01\ten{-4}$ \\
\hline\hline
\multicolumn{5}{c}{\rule{0pt}{1.2em}
$d=3$, $V_0 = {-}5.0$, $R=1.0$} \\
\hline
\rule{0pt}{1.2em}%
2 & 0.356 & $8$ & ${-}1.3(3)\ten{-2}$ & ${-}1.34\ten{-2}$ \\
3 & 2.916 & $8$ & ${-}6.2(6)\ten{-5}$ & ${-}4.80\ten{-5}$ \\
\hline\hline
\end{tabular}
\caption{Finite-volume corrections for local Gaussian potentials $V(r) = 
V_0\exp(-r^2/R^2)$ predicted from the tail of the wave function 
at a single fixed volume $L$, as described in the text.  All quantities are 
given in units of the particle mass $m=1$.}
\label{tab:Results-Gauss-DeltaB}
\end{table}

\section{Summary and outlook}

We have derived finite-volume corrections to the binding energy of a general
$N$-particle bound state in a cubic periodic volume, focusing on the leading 
behavior in this first study of such systems.  We have also derived two
new methods for computing the asymptotic normalization coefficient.  Our results 
apply to bound states with any spin and in any number of spatial dimensions, 
provided only they satisfy the condition of vanishing c.m.\ motion that is
assumed in our derivation.  In the future, it would be interesting to 
also consider the more general case of $N$-particle systems in moving frames, 
which is relevant for the scattering of composite 
particles~\cite{Bour:2011ef,Rokash:2013xda} and as tool to reduce the overall 
volume dependence of calculations (see \eg~Ref.~\cite{Davoudi:2011md}).

The results presented here should have wide applications to many 
calculations of hadronic structure, nuclear structure, and bound cold atomic 
systems.  In particular our results should be useful for lattice simulations of 
nuclei and hypernuclei starting from quarks and gluons in lattice quantum 
chromodynamics~\cite{Nicholson:2015pys,Berkowitz:2015eaa,Yamazaki:2015asa,
Yamazaki:2015vjn,Inoue:2014ipa,Etminan:2014tya,Beane:2014ora,Chang:2015qxa,
Savage:2016kon} or protons, neutrons, and hyperons in lattice effective field 
theory~\cite{Epelbaum:2013paa,Elhatisari:2015iga,Elhatisari:2016owd}.  For such
calculations performed with unphysical pion masses, one might not know \apriori 
which $A$ + $N{-}A$ system will give the leading finite-volume dependence.  In 
that case we recommend a global analysis of all the available data to determine 
a self-consistent description.

\begin{acknowledgments}
We thank Akaki Rusetsky, Ulf-G.~Meißner, and Hans-Werner Hammer for valuable
discussions.  Partial financial support was provided by the ERC Grant No.\ 
307986 STRONGINT, by the Deutsche Forschungsgesellschaft (DFG) under Grant SFB 
1245, and by the U.S.\ Department of Energy (DE-FG02-03ER41260).  Some
results reported here were obtained using computing time made available by the 
State of Hesse on the Lichtenberg High-Performance Computer at TU Darmstadt.
\end{acknowledgments}

\appendix

\section{Generator code}

\subsection{Basic principle}

We provide the Haskell code \texttt{Generator.hs} as supplementary material.  
This program solves the general problem of diagonalizing the Hamiltonian for $N$ 
nonrelativistic particles (with all equal masses) interacting via local 
potentials in a lattice-discretized periodic finite volume.  To that end, it 
automatically generates scripts for each desired case.  These scripts are meant 
to be run with GNU Octave or compatible software.

The generated scripts construct the position-space Hamiltonian as a 
(potentially very large) sparse matrix, which is then diagonalized to find the 
lowest-lying energy eigenstates.  The center-of-mass motion is removed by 
working with relative coordinates, all measured with respect to the last 
($N$-th) particle.  Periodic boundary conditions are implemented with index 
maps 
for each of these 
coordinates, and shifted versions thereof are used to generate the kinetic 
terms, which are differential operators in this representation, as finite 
differences to $k$-th order accuracy (the code supports arbitrary even 
$k\geq2$).  For further explanations regarding the code we refer to the 
comments 
included in the source file.

\subsection{Usage}
Using the generator code is very easy provided that a Haskell toolchain is 
installed on the target system; it does not require any packages beyond the 
Haskell standard library.  \texttt{Generator.hs} can either be compiled into an 
executable file or, simpler yet, be invoked directly as a script.  For example, 
the command
\begin{verbatim}
$ runhaskell Generator.hs -d 1 -n 3 -L 16
\end{verbatim}
will generate a script for three particles in one dimension, interacting via a 
two-body potential that defaults to a Gaussian well with $V_0={-}1$ and $R=1$.  
This, as well as many other parameters, can be changed with command-line 
options.  The available options that can be customized are easily identified in 
the source code.

With the above command, the generated script code is written directly to the 
terminal output.  From there, it can either be redirected to a file for further 
editing, or be conveniently passed directly to a compatible script interpreter. 
If GNU Octave is installed, a command like
\begin{verbatim}
$ runhaskell Generator.hs -d 1 -n 3 -L 16 | octave 
\end{verbatim}
will simply present the result of the numerical diagonalization to the terminal
output.

\end{document}